\documentclass[a4paper]{article}

\usepackage{INTERSPEECH2021}
\usepackage{amsmath,graphicx}
\usepackage[dvipsnames]{xcolor}
\usepackage{multirow}
\usepackage{booktabs,caption}
\usepackage[flushleft]{threeparttable}
\usepackage{hhline}

\title{Non-local Convolutional Neural Networks (NLCNN) for Speaker Recognition}

\name{$^{\star}$Haici Yang$^{\dagger}$$^{\xi}$, $^{\star}$Hongda Mao$^{\xi}$, Ruirui Li$^{\xi}$, Chelsea J.T. Ju$^{\xi}$, Oguz Elibol$^{\xi}$\thanks{*Equal contribution}}

\address{$^{\xi} $Amazon.com, Sunnyvale, USA\\
$^{\dagger}$ Indiana University Bloomington,  Bloomington, USA}
\email{$^{\dagger}$hy17@iu.edu, $^{\xi}$\{hongdam, ruirul, juitij, oelibol\}@amazon.com
}

\begin{document}

\maketitle
\begin{abstract}
Speaker recognition is the process of identifying a speaker based on their voice. The technology has attracted more attention with the recent increase in popularity of smart voice assistants, such as Amazon Alexa. In the past few years, various convolutional neural network (CNN) based speaker recognition algorithms have been proposed and achieved satisfactory performance. However, convolutional operations are building blocks that typically perform on a local neighborhood at a time and thus miss to capturing global, long-range interactions at the feature level which are critical for understanding the pattern in a speaker's voice. In this work, we propose to apply Non-local Convolutional Neural Networks (NLCNN) to improve the capability of capturing long-range dependencies at the feature level, therefore improving speaker recognition performance. Specifically, we introduce non-local blocks where the output response of a position is computed as a weighted sum of the input features at all positions. Combining non-local blocks with pre-defined CNN networks, we investigate the effectiveness of NLCNN models. Without extensive tuning, the proposed NLCNN models outperform state-of-the-art speaker recognition algorithms on the public Voxceleb dataset. What's more, we investigate different types of non-local operations applied to the frequency-time domain,  time-domain,  frequency domain and frame-level respectively. 
Among them, time domain is the most effective one for speaker recognition applications.
\end{abstract}
\noindent\textbf{Index Terms}: CNN, Speaker Recognition, Speaker Identification, Non-local Neural Networks, Self-attention

\section{Introduction}
\label{sec:intro}

Speaker recognition is the task of identifying a speaker based on 
their voice. Improving such systems has been an active area of research as there are many more practical applications, thanks to the increase 
in the popularity of smart voice assistants such as Amazon Alexa, Google Assistant, Apple Siri, and others.

Speaker recognition can be categorized into two tasks: speaker verification and speaker identification. 
Speaker verification is the task of determining whether a voice belongs to a specific speaker, while speaker identification aims at classifying the identity of an unknown voice among a set of known speakers. In addition, according to the content used for recognition, speaker recognition algorithms can also be categorized as either text-dependent or text-independent. Despite active research and significant progress in the past few years, speaker recognition remains challenging in the speech community \cite{xie2019utterance, nagrani2020voxceleb,hajavi2019deep, Li_2020_icassp,david_icassp_2018}. Traditional speaker recognition methods such as \emph{i-vector} systems are still used in the field \cite{Dehak2010,Romero2011}; however, deep neural network (DNN) based systems have become increasingly popular due to their ability to capture complex relationship among voice features and resulted in better accuracy \cite{nagrani2020voxceleb,hajavi2019deep,li2017deep,ding2020autospeech,Chung2020_odyssey,Hee2020_interspeech}.

Among the DNN-based approaches, recently, various end-to-end convolutional neural network (CNN) based speaker recognition approaches have been proposed and have achieved state-of-the-art performance \cite{xie2019utterance,chung2020defence,ding2020autospeech}. In \cite{xie2019utterance}, the authors used simplified ResNet model for frame-level feature extraction and then aggregated the features into a fixed length vector by using a dictionary-based NetVLAD or GhostVLAD for end-to-end speaker recognition. In \cite{chung2020defence}, the authors conducted a comparative study by using different CNN architectures (VGG and ResNet) and loss functions for speaker recognition on public Voxceleb dataset. Similar to \cite{xie2019utterance,chung2020defence}, most of CNN-based approaches used off-the-shelf backbone networks for feature extraction which were originally designed for computer vision tasks. In \cite{ding2020autospeech}, the authors proposed a neural architecture search (NAS) approach to automatically learn the best CNN architecture for speaker recognition with primitives again typically used in vision models.

% the drawbacks of existing approaches
Although CNN-based approaches have achieved good performance on speaker recognition, CNN operations are building blocks that are designed to operate within a local neighborhood at a time and thus miss the long-range dependency information which is critical for understanding the pattern of a speaker's voice. In speaker recognition, we hypothesize that voice features are possessed on a global scale and can be better captured by utilizing the long-range dependency information. Thus, we introduced non-local neural networks (NLCNN), which was inspired by the work~\cite{wang2018non}. The non-local operation was originally applied to image de-noising by taking into consideration the weighted mean of all the pixels in an image \cite{buades2005non}. It was then wrapped into a standalone network block and integrated within the existing network architectures~\cite{wang2018non}. 

In this work, we propose NLCNN for speaker recognition. NLCNN integrates the local information from traditional CNN layers and global, long-range information from non-local blocks. This local and global fused property of our models makes them more accurate and robust for speaker recognition. We have extensively evaluated our models on speaker verification task on the public voxceleb dataset \cite{nagrani2020voxceleb}. The results show our NLCNN models outperform state-of-the-art works.

\section{Methodology}
\label{sec:metho}

\subsection{Formulation of Non-local Operation}
\label{sec:operations}

Non-local operations were initially introduced to capture  long-range dependency information in computer vision tasks, such as object detection \cite{wang2018non}. The definition of non-local operation can be represented as the following equation:

\begin{equation}\label{equ: nl_entire}
    y_{(i,j)} = \frac{1}{C(x)}\sum_{\forall (h,k)}f(x_{(i,j)},x_{(h,k)})g(x_{(h,k)})
\end{equation}

Here, $(i,j)$ indicates the index of an output position in a 2-dimensional feature map whose response is to be computed; $(h,k)$ is the index of all possible positions within the feature map. $x$ represents the input signal and $y$ represents the output signal which has the same dimension as $x$. A pairwise function $f$ computes the relationship between samples in positions $(i,j)$ and $(h,k)$. There are many choices of $f$. In our experiment, we opted for the embedded Gaussian version introduced in \cite{wang2018non}, where $f(x_{(i,j)},x_{(h,k)}) = e^{\theta(x_{(i,j)})^T\phi(x_{(h,k)})}$. The unary function $g$ computes a representation of the input signal at the position $(h,k)$. The output response is then normalized by a factor ${C(x)}$. 

While pure convolutional operation only sums up the weighted input in a local neighborhood (e.g. $i-1\leq h\leq i+1$ and $j-1\leq k\leq j+1$ in a 3x3 kernel), non-local operation declared its non-local behavior by taking all positions $\forall (h,k)$ into the computation. Considering the properties of audio spectrograms, we also investigate applying non-local operation only on time-domain, frequency domain and frame-level. 

\textbf{Time domain}
% In the case of time-domain non-local, we assume that only pixels along the time dimension (coming from the same frequency level) have effect on the target pixel. 
When the non-local operation is only applied on time domain, the output response of a position is the weighted sum of values along the time axis. The equation (\ref{equ: nl_entire}) can be re-formulated as the following:
\begin{equation}\label{equ: time}
    y_{(i,j)} = \frac{1}{C(x)}\sum_{\forall k}f(x_{(i,j)},x_{(i,k)})g(x_{(i,k)})
\end{equation}

\textbf{Frequency domain.} Similar as non-local operation applied on the time domain, the output response of a position is the weighted sum of the positions along the frequency axis. We can get the following equation:
\begin{equation}\label{equ: frequency}
    y_{(i,j)} = \frac{1}{C(x)}\sum_{\forall h}f(x_{(i,j)},x_{(h,j)})g(x_{(h,j)})
\end{equation}

\begin{figure}[t]
  \centering
  \includegraphics[width=0.95\columnwidth]{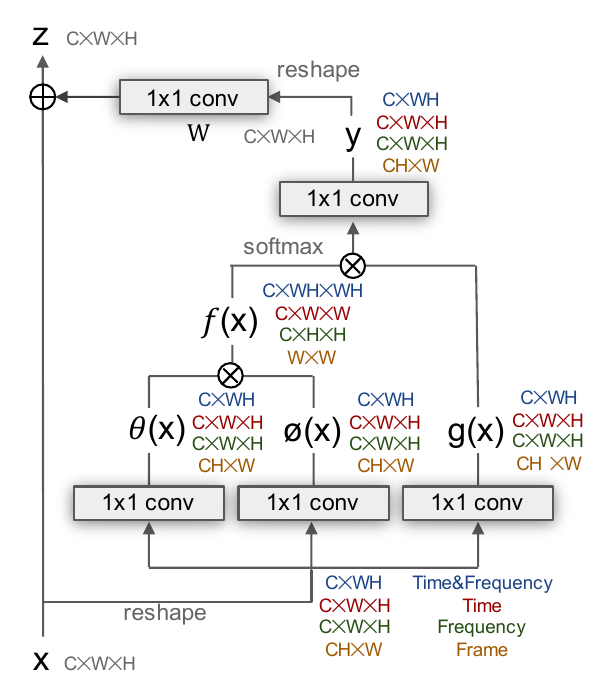}
  \caption{Non-local block with embedded Gaussian. All the linear transformation are implemented as a single convolution layer with kernel size set to 1. Four types of shape are shown for each parameter metrics, which from top to bottom associating with non-local operations on time-frequency domain, time domain, frequency domain and frame-level respectively. C, H, W represent for channal size, frequency-axis dimension and time-axis frames.
   $\oplus$  denotes element-wise addition; $\otimes$ demotes element-wise multiplication.
  }
  \label{fig:subjective}\vspace{-0.15in}
\end{figure} \label{fig:non-local}
\textbf{Frame-level.}
When we apply non-local operation on frame level, we take one frame as a group, which contains embeddings from all the frequency levels of a single time clip. As we only care about the relationship between frames, frequecy index from the same frame will be applied with the same non-local weights. The non-local weight calculated from frame level method has the smallest resolution, compared to the other pixel-wise methods discussed. However, it addressed the characteristic of sequence and simulated the way Transformer \cite{vaswani2017attention} worked on speech processing problems \cite{karita2019comparative}. Taking into account the relevant source-side information, frame level non-local operation behaves as a self-attention but in the CNN scheme.

\begin{equation}\label{equ: frame-level}
     y_{(i,j)} = \frac{1}{C(x)}\sum_{\forall k}f(x_{(:,j)},x_{(:,k)})g(x_{(:,k)})
\end{equation}

\subsection{Non-local Block}
Similar as in \cite{wang2018non}, we wrap the non-local operations in equation (\ref{equ: nl_entire}), (\ref{equ: time}), (\ref{equ: frequency}) and (\ref{equ: frame-level}) into non-local blocks which can be incorporated into any existing pre-defined CNN architectures. A non-local block can be represented as: 
\begin{equation}\label{equ: non-block}
    z_{(i,j)} = W_zy_{(i,j)} + x_{(i,j)}.
\end{equation}
Here, $y_{(i,j)}$ is given in equation (\ref{equ: nl_entire}), (\ref{equ: time}) and (\ref{equ: frequency}), and $+x_(i,j)$ indicates a residual connection which was introduced in the ResNet model \cite{He2015}. This allows us to insert a new non-local block into any pre-trained model, without breaking its initial behavior. An example of non-local block structure is illustrated in Fig. 1. The pairwise computation $f(x_{(i,j)},x_{(h,k)}) = e^{\theta(x_{(i,j)})^T\phi(x_{(h,k)})}$ can be simply done by matrix multiplication as shown in Fig. 1.

\subsection{NLCNN-based Speaker Recognition}
The non-local block is a generic, standalone network structure that can be integrated into any pre-defined CNN-based networks. To understand its behavior in speaker recognition task, we selected a recently proposed network Fast ResNet-34 for speaker recognition as our baseline model \cite{chung2020defence}. The same authors further improved the model accuracy by utilizing various data augmentation techniques and achieved state-of-the-art performance in Voxceleb dataset \cite{Hee2020_interspeech}. Since we are not using data augmentation techniques in our model training, we use \cite{chung2020defence} as our baseline model.

\textbf{Baseline model.}
 The model is a simplified ResNet-type network which is derived from the original ResNet-34 \cite{He2015}. The input feature is 40-dimensional log filterbanks energy (LFBE). Compared to the original ResNet-34 \cite{He2015}, this model used only one-quarter of the channels in each residual block in order to reduce computational cost. It also removed the maxpooling layer to avoid downsampling the feature map too early.  Detailed model architecture is shown in Table \ref{table-baseline_model}.
 
 \textbf{NLCNN model.} We inserted non-local blocks into the baseline model and turned it into a NLCNN model. Based on different number  and  location of non-local blocks inserted, we generated a series of NLCNN models. More details can be found in Section~\ref{sec:experiments}.  

\begin{table}[t]

% \fontsize{5}{5}
\centering
\setlength{\belowcaptionskip}{10pt}
\resizebox{0.95\linewidth}{!}{
\begin{tabular}{c|c|c}
\hline
 & Layer & Output size                      \\ \hline \hline
conv1 &  3$\times$3, 16, stride=(2,1)  & 16$\times$20$\times$ T \\ \hline
%1 in Res1*, 2 in Res2
conv2\_x&   [
                \begin{tabular}{ll}
                  3$\times$3, 16\\
                  3$\times$3, 16
                \end{tabular}
              ] $\times$3,  stride (1,1)
  & 16$\times$20$\times$ T \\ \hline
%1 in Res2, 2 in Res3 
conv3\_x&  [
                \begin{tabular}{ll}
                  3$\times$3, 32\\
                  3$\times$3, 32
                \end{tabular}
              ]$\times$3,  stride (2,2)
  & 32$\times$10$\times$ T/2 \\ \hline
%1 in Res3, 2 in Res4 
conv4\_x&   [
                \begin{tabular}{ll}
                  3$\times$3, 64\\
                  3$\times$3, 64
                \end{tabular}
              ]$\times$3,  stride (2,2)
  & 64$\times$5$\times$ T/4 \\ \hline
   
conv5\_x&   [
                \begin{tabular}{ll}
                  3$\times$3, 128\\
                  3$\times$3, 128
                \end{tabular}
              ]$\times$3,  stride (1,1)
  & 128$\times$5$\times$ T/4 \\ \hline
aggregation & SAP & 128$\times$1 \\ \hline
fc & 128$\times$512 & 512$\times$1 \\ \hline

\end{tabular}
}
\caption{ \small {
Our baseline model Fast ResNet-34 for speaker recognition \cite{chung2020defence}. The dimensions of the output feature map size is  $channels \times height \times width$. The input size is $1 \times 40 \times T$, where $T$ is the number of frames of the input.  The residual blocks are shown in brackets.
}}
\label{table-baseline_model}
% \vspace{-8mm}
\end{table}

\section{Experiments}
Like other end-to-end neural network based speaker recognition systems \cite{chung2020defence,xie2019utterance,wan2018generalized}, our pipeline includes four major components: (1) Input representation. We used a 40-dimensional log filterbank energy (LFBE) which was extracted from the raw audio with a hamming window 25ms and 10ms shift. (2) Frame-level feature extraction. Either NLCNN models or baseline models are used to extract frame-level feature. (3) Utterance-level feature aggregation. We tested both Self-attentive pooling (SAP) \cite{cai2018exploring} and GhostVLAD \cite{zhong2018ghostvlad}. As the two methods perform similarly in our experiments, we only present the results with the SAP approach. (4) Loss function for network optimization. We utilized two different types of loss function for model optimization: AM-Softmax (AMS) and Angular Prototype (AP). AMS is a classification-based loss function, which aims at explicitly mapping the query speaker to an exact speaker profile in the data set. AP behaves as verification-based loss function, merely working to differentiate N speakers appearing within a batch \cite{chung2020defence}. 

\begin{table}[t]
\tiny
\centering
\resizebox{0.95\linewidth}{!}{
\begin{tabular}{c|c|c|c}
\hline
 & AMS & AP &  \# Parameters    \\ \hhline{====}
Baseline-1 \cite{chung2020defence}  & 2.43 $\pm$ $0.05$ & 2.22 $\pm$ $0.05$ & 1.40M \\ 
Baseline-3 &  2.46 $\pm$ $0.04$& 2.23 $\pm$ $0.05$& 1.44M\\ 
Baseline-6  & 2.35 $\pm$ $0.03$ & 2.37 $\pm$ $0.04$ & 1.46M\\ \hline
1-block NLCNN  & 2.44 $\pm$ $0.03$ & 2.22 $\pm$ $0.04$ & 1.40M\\ 
3-block NLCNN  & \textbf{2.22} $\pm$  $0.03$   & \textbf{2.08} $\pm$ $0.06$ & 1.44M\\  
6-block NLCNN  & 2.25 $\pm$ $0.02$ &  2.19 $\pm$ $0.02$  & 1.46M\\ \hline
\end{tabular}
}
\caption{\small{The effect of number of non-local blocks on test performance. We compare NLCNN models with 1, 3 and 6 non-local blocks to their corresponding baseline models. We evaluated models trained with two different loss functions: AMS and AP. We reported mean and standard deviation of repeated experiments.}}
\label{tab:diff_layers}
% \vspace{-10mm}
\end{table}

\begin{table}[t]
\tiny
\centering
\setlength{\belowcaptionskip}{10pt}%
\resizebox{0.95\linewidth}{!}{

\begin{tabular}{c|c|c|c}
\hline
 Models& AMS & AP   & \# Parameters                    \\ \hhline{====}
Baseline-3 &  2.46  $\pm$ $0.04$  & 2.23  $\pm$ $0.05$& 1.44M \\ \hline
%1 in Res1*, 2 in Res2
Var1& \textbf{2.22}  $\pm$ {$0.08$} & 2.09  $\pm$ $0.06$  & 1.44M\\ 
%1 in Res2, 2 in Res3 
Var2& 2.26  $\pm$ {$0.06$} & \textbf{2.07}  $\pm$ $0.06$ & 1.44M \\  
%1 in Res3, 2 in Res4 
Var3& 2.27  $\pm$ {$0.03$} & 2.23  $\pm$ $0.02$ &1.44M\\ \hline

\end{tabular}

}
\caption{\small{The effect of locations of non-local blocks on test performance. 3 non-local blocks are inserted into different locations of the baseline model. 'Var1' represents the NLCNN model with 1 non-local block inserted at conv2\_x and 2 non-local blocks inserted at conv3\_x; 'Var2' represents the NLCNN model with 1 block inserted at conv3\_x and 2 non-local blocks inserted at conv4\_x; 'Var3' represents the NLCNN model with 1 block inserted at conv4\_x and 2 non-local blocks inserted at conv5\_x. We reported mean and standard deviation of repeated experiments.}}
\label{tab:diff_location}
% \vspace{-6mm}
\end{table}

\begin{table}[t]
% \tiny
\centering
\setlength{\belowcaptionskip}{10pt}%
\resizebox{0.9\linewidth}{!} {
\begin{tabular}{c|c|c|c|c}
\hline
Variants &Time+Frequency & Time  & Frequency & Frame     \\ \hhline{=====}

 Var1 & 2.42  $\pm$ $0.02$ &\textbf{2.18} $\pm$ $0.07$ &2.26  $\pm$ $0.02$&2.33  $\pm$ $0.05$ \\ 
Var2 & 2.28  $\pm$ $0.06$ &\textbf{2.23}  $\pm$ $0.03$ &2.49  $\pm$ $0.02$ &2.32  $\pm$ $0.01$ \\ 
 Var3 &{2.26}  $\pm$ $0.04$ &\textbf{2.29}  $\pm$ $0.02$ &2.51  $\pm$ $0.03$ & 2.33  $\pm$ $0.04$\\ \hline
% \multicolumn{2}{c|}{mel-40 6-block}& 2.26&\textbf{2.23}&2.64&2.42\\ \hline
% \multicolumn{2}{|c|}{stft-257 1-block}&2.14&\textbf{1.96}&2.87 &2.45\\ \hline
\end{tabular}
}
\caption{\small{The effect of types of non-local operations on test performance. The NLCNN model architecture of Var1, Var2 and Var3 are the same as described in Table \ref{tab:diff_location}. Only the results of models trained with AMS are listed here. We reported mean and standard deviation of repeated experiments.}}
\label{table-label-2}
% \vspace{-6mm}
\end{table}

\label{sec:experiments}

\subsection{Datasets}
In the experiments, we trained our models with Voxceleb2 \cite{chung2018voxceleb2} dev dataset which contains over 1 million utterances from 5,994 speakers. We then evaluated the models on Voxceleb1 \cite{nagrani2017voxceleb} test set which contains 4872 utterances produced by 40 speakers. There is no speaker overlap between Voxceleb2 dev set and Voxceleb1 test set. Since  Voxceleb dataset consists mostly continuous speech, neither voice activity detection (VAD) nor automatic silence removal is applied.

\subsection{Model training}
Our pipeline was implemented in Pytorch and was trained on AWS EC2 machine with a single NVIDIA Tesla V100 GPU. All the models were trained for 500 epochs, and the best checkpoint was selected for model evaluation. During model training, we randomly select 2-second segments from the utterances to form mini-batches, and ensure that utterances from the same speaker should be less than 100 in each training epoch. 
% This training strategy is in line with \cite{chung2020defence}.
We used Adam optimizer for optimization with learning rate starts at 0.001 and decreases by 5\% every 10 epochs. No data augmentation was used during model training. 

\subsection{Model evaluation}
We evaluated the models on speaker verification task with Equal Error Rate (EER) which is the value when the false-reject rate equals the false-accept rate. To have a fair comparison with our baseline models \cite{chung2020defence}, we used the same evaluation protocol by sampling ten 4-second temporal crops at regular intervals from each test utterance, and the mean similarity score of the all possible pairs are used as the final score. 
% As we used 2-second segment for training, we also evaluated the models with 2-second segment with the same setting. 
This evaluation protocol is in line with \cite{chung2020defence,Chung2020_odyssey,chung2018voxceleb2}. We repeated the model training with the same architecture setting for three times, and report the mean and standard deviation of the repeated experiments.

\subsection{Results}

\subsubsection{The effect of non-local blocks}
We first investigated the performance of NLCNN models with different number of non-local blocks. Specifically, we added 1, 3 and 6 non-local blocks to the baseline model and generated the corresponding NLCNN models. To make a fair comparison, we increased the depth of the baseline model by purely inserting CNN layers. As there are no existing performance numbers available for Baseline-3 and Baseline-6, we trained and evaluated the models based on the authors' released code \cite{chung2020defence}. 

 As shown in Table \ref{tab:diff_layers}, both 3-block and 6-block NLCNN models achieved consistently lower EER than the corresponding baseline models with same amount of parameters. It is noteworthy that the performance of both baselines and NLCNN models are not simply improved by stacking more convolutional layers or non-local blocks. While 1-block NLCNN barely gives the original baseline model a boost, 3-block NLCNN model significantly reduces the EER by over $10\%$. With two more non-local block stacked, 3-block NLCNN apparently takes advantage of the accumulated influence of non-local operation as opposed to 1-block NLCNN. However, the effect is not unbounded. 6-block NLCNN only marginally improves the performance.

In general, we are confident to argue that moderate amount of global, long-range interaction information extracted by non-local blocks efficiently improves the overall speaker verification accuracy, since global voice pattern is critical for distinguishing one voice from another.
Furthermore, models trained with AP loss function generally performed better than models trained with AMS under the same setting. This observation is consistent with the findings in \cite{chung2020defence}.

\subsubsection{The effect of different locations of non-local blocks}
In Table \ref{tab:diff_location}, we showed the performance of three different NLCNN models where three non-local blocks insert at different locations in the baseline model. Although there is some performance variance among the models, all the NLCNN models performed consistently better than the baseline model.  More importantly, Var1 and Var2 on average give better results in contrast with Var3. One explanation could be that the feature maps size at the top layers, i.e. conv4\_x and conv5\_x, is already quite small and each position has represented a considerably large area of information at high-level, as a result of convolution and striding, thus leaving limited space for the further non-local information learning, and blunts the model's response to non-local blocks. On the contrary, with smaller span of information represented in each location of the feature map, the non-local weight maps of those bottom convolutional layers, i.e. conv2\_x and conv3\_x, become intricate, and reward the aid of non-local blocks.

\subsubsection{The effect of types of non-local operation}

We also investigated the effect of four different types of non-local operations discussed in section \ref{sec:operations}. As shown in
Table \ref{table-label-2}, NLCNN models with time-domain non-local operation achieved the lowest EER among all the models. This finding is slightly different from the work \cite{wang2018non} with computer vision tasks where frequency-time (space-time) operation achieved the best performance. We believe this is because global time-domain interaction information is more important for sequence data, while the translation invariant property of image data makes space-time more important for vision tasks. 

It is also interesting to notice that non-local models in frame-level are mediocre although they also learns non-local knowledge horizontally through the feature maps as what the time-level models do. Our non-local operation becomes self-attention-like form when it operates at frame level as formulated in equation (\ref{equ: frame-level}), that is positions belong to the same frame will receive identical non-local weights. That being the case, the performance discrepancy between time-level and frame-level non-local model uncovers the fact that the weights learned from non-local operations are essentially frequency-variant in the speaker recognition problem. Although the model benefit more from time-level neighbours than from frequency-level neighbours, each frequency row requires to preserve its unique pattern of time-level non-local distribution. The time+frequency non-local model does include frequency-wise calculation, however, the excessively high level of non-local granularity clearly burdens the model and deteriorates the performance thereof. Based on the result, we conclude that non-local operation along the time domain works the best for speaker recognition task. Furthermore, our results suggest the substance of retaining frequency-wise operations in attentional model, especially for speaker recognition applications.

\section{Conclusion}
\label{sec:conclusion}

In this work, we presented NLCNN for speaker recognition.  NLCNN can capture global, long-range dependency information at feature level via non-local blocks. 
Combined with pre-defined CNN-based neural network, we investigated the effect of non-local blocks on speaker recognition with different types of operations and different  positions in the network.  We found out that time domain is the most effective one for speaker recognition application. Our experiments also showed that NLCNN models outperforms state-of-the-art works on speaker recognition. 

\newpage

\bibliographystyle{IEEEtran}
\bibliography{mybib}

\end{document}